\documentclass[11pt]{article}
\usepackage{graphicx}
\usepackage{amssymb}
\usepackage{epsfig}

\newtheorem{proposition}{Proposition}

\def\proof{\noindent  {\underline {Proof}}. }
\def\square{ {\hfill \vrule height6pt width6pt depth1pt} \bigskip \medskip }

\def\proof{\noindent  {\underline {Proof}}. }
\def\square{ {\hfill \vrule height6pt width6pt depth1pt} \newline  }

\def\today{\number\day\space\ifcase\month\or Janvier \or F\'evrier \or  Mars
   \or Avril \or Mai \or Juin \or Juillet \or Ao\^ut \or Septembre \or Octobre
   \or Novembre \or D\'ecembre \fi\number \year}

\def\bea{\begin{eqnarray}}
\def\eea{\end{eqnarray}}

\begin{document}
\hfill ITEP-TH-07/07 \vskip 2.5cm
\centerline{On the Bethe Ansatz for the Jaynes-Cummings-Gaudin model}
\bigskip
\centerline{O. Babelon\footnote{
Laboratoire de Physique Th\'eorique et Hautes Energies (LPTHE), Tour 24-25, 5\` eme \'etage, Boite 126,
4 Place Jussieu,  75252 Paris Cedex 05,
Unit\'e Mixte de Recherche UMR 7589,
Universit\'e Pierre et Marie Curie-Paris6; CNRS; Universit\'e Denis Diderot-Paris7.
}
, D. Talalaev\footnote{Institute for Theoretical and Experimental Physics (ITEP) 117218 Russia, Moscow, B. Cheremushkinskaja 25 
}}

\bigskip
{\bf Abstract.} We investigate the quantum Jaynes-Cummings model -
a particular case of the Gaudin model with one of the spins being infinite.
Starting from the Bethe equations
we derive Baxter's equation and from it a closed set of equations for the eigenvalues of the commuting Hamiltonians.
A  scalar product in the separated variables representation is found for which the commuting Hamiltonians are Hermitian. 
In the semi classical limit the Bethe roots accumulate on  very specific  curves in the complex plane.  We give the equation of these curves. They build up a system of cuts  
modeling  the spectral curve as a two sheeted cover of the complex plane. 
Finally, we  extend some of these results to the XXX Heisenberg spin chain.

\section{Introduction}

The Jaynes-Cummings-Gaudin model is defined by the Hamiltonian
\begin{equation}
H = \sum_{j=0}^{n-1} 2 \epsilon_j s_j^z + \omega b^\dag b  + g \sum_{j=0}^{n-1} (b^\dag s_j^- + b s_j^+)
\label{JCGHam}
\end{equation}
Here $b, b^\dag$ is a quantum harmonic oscillator 
$$
[ b , b^\dag ] = \hbar 
$$
and $s_j^z, s_j^\pm$ are quantum spin operators
$$
 [ s_j^+ , s_j^- ] = 2\hbar s_j^z,\quad  [ s_j^z , s_j^\pm ] = \pm\hbar s_j^\pm
$$
For the oscillator, we represent $b,b^\dag$ as
\begin{equation}
b = \hbar {d\over d z}, \quad b^\dag = z
\label{bargplat}
\end{equation}
They act on the Bargman space
$
{\cal B}_b = \left\{ f(z),{\rm entire~function~of~} z ~\Big \vert \int  \vert f(z) \vert^2 e^{- {\vert z \vert^2 \over \hbar }}  dz d\bar{z} < \infty \right\}
$.
For the spin operators, we assume that $s_j^a$ acts on a spin $s_j$ representation
\begin{eqnarray*}
s_j^z \vert m_j \rangle &=& \hbar m_j \vert m_j \rangle \\
s_j^\pm \vert m_j \rangle &=& \hbar  \sqrt{s_j(s_j+1) - m_j(m_j \pm 1)}  \; \vert m_j \pm 1\rangle,
\quad m_j = -s_j, -s_j+1, \cdots , s_j-1,s_j
\end{eqnarray*}
where $s_j$ is integer or half integer. 

The Jaynes-Cummings model is well known in condensed matter physics \cite{JC,AR,NSE}. It also appears in the book by M. Gaudin \cite{Gau83},  but the connection between the two seems to be recent \cite{YKA}.

\section{Bethe Ansatz}
In order to write the Bethe Ansatz, we introduce the Lax matrix
$$
L(\lambda) = \pmatrix{ A(\lambda) & B(\lambda) \cr C(\lambda) & -A(\lambda)}
$$
where the operator valued matrix elements are defined as
\begin{eqnarray*}
A(\lambda) &=& {2\lambda \over g^2} -{\omega \over g^2} + \sum_{j=0}^{n-1} {s^z_j\over \lambda - \epsilon_j} \\
B(\lambda) &=& {2 b \over g} + \sum_{j=0}^{n-1} {s^-_j\over \lambda - \epsilon_j} \\
C(\lambda) &=& {2 b^\dag \over g} + \sum_{j=0}^{n-1} {s^+_j\over \lambda - \epsilon_j}
\end{eqnarray*}
It is simple to check the commutation relations
\begin{eqnarray*}
~ [A(\lambda), B(\mu) ] &=& {\hbar \over \lambda - \mu} ( B(\lambda) - B(\mu) ) \\
~ [A(\lambda), C(\mu) ] &=& -{\hbar \over \lambda - \mu} ( C(\lambda) - C(\mu) ) \\
 ~[B(\lambda), C(\mu) ] &=& {2 \hbar \over \lambda - \mu} ( A(\lambda) - A(\mu) )
\end{eqnarray*}
Moreover one has
$ [A(\lambda), A(\mu) ] = 0$, $  [B(\lambda), B(\mu) ] =0$ and  $[C(\lambda), C(\mu) ] =0$.
Defining
$$
{1\over 2}{ \rm Tr}\,(L^2(\lambda) ) =  A^2(\lambda) +{1\over 2}( B(\lambda) C(\lambda) + C(\lambda) B(\lambda))
$$
We have
$
[ { \rm Tr}\, L^2(\lambda)  , {\rm Tr}\, L^2(\mu) ] = 0
$
so that ${\rm Tr}\, L^2(\lambda)$ generates a family of commuting quantities. Expanding in $\lambda$ we get
\begin{equation}
{1\over 2}{ \rm Tr}\,(L^2(\lambda) )  =  {1\over g^4}(2\lambda - \omega)^2 + {4\over g^2} H_n 
+ {2\over g^2}\sum_{j=0}^{n-1} {H_j\over \lambda - \epsilon_j} + \sum_{j=0}^{n-1} {\hbar^2 s_j(s_j+1) \over (\lambda-\epsilon_j)^2}
\label{defHj}
\end{equation}
The Hamiltonian eq.(\ref{JCGHam}) is given by
$$
H = \omega H_n + \sum_{j} H_j
$$

To write the Bethe Ansatz we define the reference state which is the lowest weight vector:
$$
\vert 0 \rangle = \vert 0\rangle \otimes \vert -s_1 \rangle \otimes \cdots \otimes \vert -s_n \rangle,
 \quad b  \vert 0 \rangle = 0, \quad s_j^-  \vert  -s_j \rangle = 0
$$
This vector has the following important properties
$$
B(\lambda) \vert 0 \rangle = 0
$$
and
$$
A(\lambda) \vert 0 \rangle = a(\lambda) \vert 0 \rangle, \quad a(\lambda) = {2\lambda\over g^2}
-{\omega \over g^2} -\sum_j { \hbar s_j\over \lambda-\epsilon_j}
$$
Moreover since $[B(\lambda) , C(\lambda) ] = 2 \hbar A'(\lambda)$
 we also have
$$
B(\lambda) C(\lambda) \vert 0 \rangle = 2 \hbar  a'(\lambda) \vert 0 \rangle
$$
With all this we deduce
$$
{1\over 2} {\rm Tr}\, L^2(\lambda)  \vert 0 \rangle = (a^2(\lambda) + \hbar a'(\lambda) )  \vert 0 \rangle
$$
Let us now define the vector
$$
\Omega(\mu_1, \mu_2, \cdots, \mu_M) = C(\mu_1) C(\mu_2) \cdots C(\mu_M) \vert 0 \rangle
$$
It is not difficult to prove that  (see e.g. \cite{Gau83})
$$
{1\over 2} {\rm Tr}\, L^2(\lambda)  \Omega(\mu_1, \mu_2, \cdots, \mu_M)= \Lambda(\lambda, \mu_1, \mu_2, \cdots, \mu_M)\Omega(\mu_1, \mu_2, \cdots, \mu_M)
$$
\begin{equation}
\Lambda(\lambda, \mu_1, \mu_2, \cdots, \mu_M)= a^2(\lambda) + \hbar a'(\lambda) + 2\hbar \sum_i  {a(\lambda) - a(\mu_i)\over \lambda - \mu_i}  
\label{betheeigenval}
\end{equation}
provided the parameters $\mu_i$ satisfy the set of Bethe equations.
\begin{equation}
a(\mu_i)  +\sum_{j \neq  i}{\hbar \over \mu_i - \mu_j} =0
\label{bethe}
\end{equation}

\section{Riccati equation}

We now analyse the Bethe equations eqs.(\ref{bethe}). We introduce the function
$$
S(z) = \sum_i {1\over z-\mu_i}
$$

\begin{proposition}
The Bethe equations (\ref{bethe}) imply the following Riccati equation on $S(z)$
\begin{equation}
S'(z) + S^2(z) + {2\over \hbar g^2} \Big( (2z-\omega) S(z) - 2 M  \Big) =  \sum_j 2s_j { S(z) - S(\epsilon_j) \over z-\epsilon_j}
\label{ricati}
\end{equation}
\end{proposition}
\proof
The Bethe equations read
$$
{2\mu_i \over g^2}
-{\omega \over g^2} -\sum_j { \hbar s_j\over \mu_i-\epsilon_j}
 +\sum_{j \neq  i}{\hbar \over \mu_i - \mu_j} =0
 $$
 we multiply by $1/(z-\mu_i)$ to get
 $$
{2\over g^2} { \mu_i  \over z-\mu_i}
-{\omega \over g^2}{ 1  \over z-\mu_i} -\sum_j { \hbar s_j\over \mu_i-\epsilon_j}{ 1  \over z-\mu_i}
 +\sum_{j \neq  i}{\hbar \over \mu_i - \mu_j}{ 1  \over z-\mu_i} =0
 $$
We now sum over $i$. We have
$$
\sum_{i=1}^M   { \mu_i  \over z-\mu_i} = \sum_i     { \mu_i - z  \over z-\mu_i} +  {  z  \over z-\mu_i}
= -M + zS(z)
$$
also
$$
\sum_{i=1}^M {1\over \mu_i-\epsilon_j}{ 1  \over z-\mu_i} =
{1\over z-\epsilon_j} \sum_i \left({1\over z-\mu_i} + {1\over \mu_i-\epsilon_j}\right) =
{S(z)-S(\epsilon_j)\over z-\epsilon_j}
$$
and finally 
\begin{eqnarray*}
\sum_{i=1}^M \sum_{j\neq i} {1 \over \mu_i - \mu_j}{ 1  \over z-\mu_i}&=&
{1\over 2}\sum_{i=1}^M \sum_{j\neq i} {1 \over \mu_i - \mu_j}\left( { 1  \over z-\mu_i}
- {1\over z-\mu_j} \right) 
= {1\over 2}\sum_{i=1}^M \sum_{j\neq i}   { 1  \over (z-\mu_i)(z-\mu_j)} \\
&=& {1\over 2}\left( \sum_{i,j}  { 1  \over (z-\mu_i)(z-\mu_j)} - \sum_i  { 1  \over (z-\mu_i)^2} \right) 
=  {1\over 2} (S^2(z) + S'(z) )
\end{eqnarray*}
\square

In equation (\ref{ricati}) the $S(\epsilon_j)$ appear as parameters. The Riccati equation itself determines them as we now see.
Suppose first that $s_j=1/2$. We let $z\to \epsilon_i$ into eq.(\ref{ricati}) getting
$$
S'(\epsilon_i) + S^2(\epsilon_i) + {2\over \hbar g^2} \Big( (2\epsilon_i-\omega) S(\epsilon_i) - 2 M  \Big) =  S'(\epsilon_i) + \sum_{j\neq i}  { S(\epsilon_i) - S(\epsilon_j) \over \epsilon_i-\epsilon_j}
$$
The remarkable thing is that $S'(\epsilon_i) $ cancel in this equation and we get a
set of closed algebraic equations for the $S(\epsilon_j)$.
\begin{equation}
 S^2(\epsilon_i) + {2\over \hbar g^2} \Big( (2\epsilon_i-\omega) S(\epsilon_i) - 2 M  \Big) =
  \sum_{j\neq i}  { S(\epsilon_i) - S(\epsilon_j) \over \epsilon_i-\epsilon_j} , \quad
  i=1,\cdots, n
  \label{charac1/2}
\end{equation}

Suppose next that $s_j=1$. We expand the Riccati equation around $z=\epsilon_i$:
\begin{eqnarray*}
(z-\epsilon_i)^0&: & S'(\epsilon_i) + S^2(\epsilon_i) + {2\over \hbar g^2} ((2\epsilon_i-\omega) S(\epsilon_i) - 2M) = 2S'(\epsilon_i) + 2 \sum_{j\neq i} {S(\epsilon_i) - S(\epsilon_j)\over \epsilon_i
-\epsilon_j} \\
(z-\epsilon_i)^1&: & S''(\epsilon_i) + 2 S(\epsilon_i) S'(\epsilon_i)+ {2\over \hbar g^2} ((2\epsilon_i-\omega) S'(\epsilon_i) +2 S(\epsilon_i)) \\
&& \hskip 6cm = S''(\epsilon_i) - 2 \sum_{j\neq i} {S(\epsilon_i) - S(\epsilon_j)\over (\epsilon_i-\epsilon_j)^2} -  {S'(\epsilon_i) \over \epsilon_i-\epsilon_j}
\end{eqnarray*}
We see that in the second equation $S''(\epsilon_i)$ cancel. The first equation allows to compute
$S'(\epsilon_i)$ and the second equation then gives a set of closed equations for the $S(\epsilon_i)$.
The general mechanism is clear. For a spin $s$, we expand
$$
S'(z) - 2s {S(z)-S(\epsilon) \over z-\epsilon} = \sum_m {m-2s\over m!} S^{(m)}(\epsilon) (z-\epsilon)^{m-1}
$$
and we see that the coefficient of $S^{(2s)}(\epsilon)$ vanishes in the term $m=2s$. The equations coming from $(z-\epsilon)^{m-1}$
for $m=1, \cdots, 2s-1$ allow to compute $S'(\epsilon), \cdots, S^{(2s-1)}(\epsilon)$
by solving at each stage a linear equation. Plugging into the equation for $m=2s$ , we obtain a closed
equation of degree $2s+1$ for $S(\epsilon)$.
\begin{equation}
P_{2s+1}(S(\epsilon)) = 0
\label{E2s+1}
\end{equation}
Notice that if $M < 2s$, the system will truncate at level $M$ because there always exists a relation of the form $S^{(M)} = P(S,S',\cdots S^{(M-1)} )$.

The $S(\epsilon_j)$ also determine the eigenvalues of the commuting Hamiltonians. Going back to eq.(\ref{betheeigenval}), we see that
\begin{eqnarray*}
{a(\lambda) - a(\mu_i)\over \lambda - \mu_i}   &=& {2\over g^2} -
\sum_j { \hbar s_j \over \lambda-\mu_i}\left( {1\over \lambda-\epsilon_j}-  {1\over \mu_i-\epsilon_j} \right)
\\
&=& {2\over g^2} +
\sum_j { \hbar s_j \over \lambda-\epsilon_j} {1\over \mu_i-\epsilon_j}=  {2M \over g^2}
- \sum_j { \hbar s_j S(\epsilon_j)\over \lambda-\epsilon_j}
\end{eqnarray*}
Hence
\begin{equation}
 \Lambda(\lambda)=  a^2(\lambda) + \hbar a'(\lambda) + 2\hbar\left(
{2M \over g^2}
- \sum_j { \hbar s_j S(\epsilon_j)\over \lambda-\epsilon_j}
\right)
\label{Lambda}
\end{equation}
Expanding $a(\lambda)$ we deduce the eigenvalues of the commuting Hamiltonians
$$
h_n = \hbar M - \hbar \sum_j s_j + {\hbar\over 2},
\quad
h_j = {2\omega \over g^2} \hbar s_j -  {4\over g^2} s_j \epsilon_j - 2 \hbar^2 s_j S(\epsilon_j) 
+ 2 \sum_{i\neq j} {\hbar^2 s_j  s_i \over \epsilon_j -\epsilon_i}
$$
The algebraic equations for the $S(\epsilon_j)$ are therefore the characteristic equations of the set of matrices $H_j$. The existence of such characteristic equations is a general phenomenon. In the Appendix we derive them for the Heisenberg XXX spin chain.

\section{Baxter equation}

We can linearize the Riccati equation eq.(\ref{ricati}) by setting
\begin{equation}
S(z) = {\psi'(z)\over \psi(z)}
\label{s-and-psi}
\end{equation}
Obviously
\begin{equation}
\psi(z) = \prod_{i=1}^M (z-\mu_i)
\label{defpsi}
\end{equation}
The linearized equation reads
\begin{equation}
\psi''(z) +{2\over \hbar} a(z) \psi'(z) + {2\over \hbar} \left( -{2M\over g^2} + \sum_j{ \hbar s_j  S(\epsilon_j)
\over z-\epsilon_j} \right)\psi(z) = 0
\label{eqpsi}
\end{equation}
Here, we should understand that the $S(\epsilon_j)$ are determined  by the procedure explained
in the previous section, for instance by eq.(\ref{charac1/2}) for spins $s_j=1/2$.
For such values of the parameters, the equation has the following
remarkable property.

\begin{proposition}[Mukhin, Tatasov, Varchenko \cite{MTV}]
For the special  values of the parameters $S(\epsilon_j)$ coming from the Bethe equations,
the solutions of eq.(\ref{eqpsi}) have trivial monodromy.
\end{proposition}
\proof
Strictly speaking  the proof in \cite{MTV} is valid for the finite-dimensional representations.  We provide here a straightforward proof in our case.
The proposition is clear for the solution $\psi_1(z)$ defined by eq.(\ref{defpsi}) since it is a polynomial. A second solution can be constructed as usual
$$
\psi_2(z) = \psi_1(z) \int^z  \exp \left(-{2\over \hbar}\int^y a(t) dt  - 2 \log \psi_1(y)  \right)   dy
=  \psi_1(z) \int^z  {\prod_j (y-\epsilon_j)^{2s} \over \prod_i (y-\mu_i)^2} e^{-{2\over \hbar g^2} (y^2-\omega y)} dy
$$
The monodromy will be trivial if  the pole at $y=\mu_i$ has no residue preventing the apparition of
logarithms. Expanding around $\mu_i$, we have
\begin{eqnarray*}
\exp \left(-{2\over \hbar}\int^y a(t) dt  - 2 \log \psi_1(y)  \right)&=&
{e^{\left(-{2\over \hbar}\int^{\mu_i }a(t) dt -2 \sum_{j\neq i} (\mu_i-\mu_j)  \right)}
\over (y-\mu_i)^2}\times \\
&& \exp \left(-{2\over \hbar}(y-\mu_i) \left[a(\mu_i) + \sum_{j\neq i}{\hbar \over \mu_i-\mu_j}\right] +
O(y-\mu_i)^2 \right)
\end{eqnarray*}
but the coefficient of the dangerous $(y-\mu_i)$ term vanishes by virtue of the Bethe equations.
\square

Next we set
\begin{equation}
\psi(z) = \exp\left(-{1\over \hbar} \int^z a(y)dy \right) Q(z)
\label{psi-and-q}
\end{equation}
We obtain for $Q(z)$ the equation
$$
\hbar^2 Q''(z) - \left( a^2(z) + \hbar a'(z) + {4\hbar M\over g^2} - 2 \hbar^2 \sum_j {s_j S(\epsilon_j) \over z-\epsilon_j}\right) Q(z)
$$
Comparing with eq.(\ref{Lambda}), this is also
\begin{equation}
\hbar^2 Q''(z) -  \Lambda(z) Q(z)
\label{baxter}
\end{equation}
Hence, we have recovered Baxter's equation. Notice that
\begin{equation}
Q(z) ={e^{{1\over \hbar g^2} (z^2-\omega z) } \over \prod_j (z-\epsilon_j)^{s_j}}\psi(z) = {e^{{1\over \hbar g^2} (z^2-\omega z) } \over \prod_j (z-\epsilon_j)^{s_j}} \prod_{i=1}^M (z-\mu_i)
\label{Qpsi}
\end{equation}

\section{Bethe eigenvectors and separated variables}

We recall the form of the Bethe eigenvectors.
$$
 \Omega(\mu_1, \mu_2, \cdots, \mu_M) = C(\mu_1)\cdots C(\mu_M) \vert 0 \rangle
$$ 
 Following Sklyanin, \cite{Sk79}, we introduce the set of zeros of $C(\lambda)$
$$
C(\lambda) = {2z\over g} {\prod_{i=1}^n (\lambda - \lambda_i) \over \prod_{j=1}^n (\lambda - \epsilon_j)}
$$
The operators $\lambda_i$ commute among themselves. Inserting this expression for $C(\lambda)$
 into the Bethe state and remembering eq.(\ref{defpsi}), we find
\begin{equation}
 \Omega(\mu_1, \mu_2, \cdots, \mu_M)  = \left( \prod_j  {1\over \psi(\epsilon_j) } \right) \left({2z\over g}\right)^M
 \prod_i \psi(\lambda_i)  \vert 0 \rangle
 \label{eigensepar}
 \end{equation}
 If we  now switch to a Schroedinger representation where the $\lambda_i$ are represented 
 as multiplication operators, the eigenstate eq.(\ref{eigensepar}) is represented by a product of functions in one variable
 $\psi(\lambda_i)$. This means that we have separated the variables in the Schroedinger equation.
 \begin{proposition}
In the separated variables, the Hamiltonians read
\begin{equation}
H_j = { \prod_k (\epsilon_j - \lambda_k)\over \prod_{k\neq j}(\epsilon_j - \epsilon_k)}
 \sum_i { \prod_{k\neq j} (\lambda_i-\epsilon_k) \over \prod_{k\neq i} (\lambda_i-\lambda_k) } \left( {d^2\over d\lambda_i^2} +{2\over \hbar} a(\lambda_i) {d\over d\lambda_i} -{M\over \hbar}  \right)
 \label{hamsepa}
\end{equation}
\end{proposition}
\proof
Let us introduce the set of commuting operators $H_j$ diagonal in the Bethe states basis
eq.(\ref{eigensepar}), (here we assume completeness of Bethe Ansatz), and  such that
$$
H_j \prod_k \psi(\lambda_k) =  2 s_j S(\epsilon_j) \prod_k \psi(\lambda_k) 
$$
These are essentially the same operators as in eq.(\ref{defHj}). Then eq.(\ref{eqpsi}) implies for each variable $\lambda_i$
$$
\sum_j {1\over \lambda_i - \epsilon_j} H_j \prod_k  \psi(\lambda_k) =
- \left( {d^2\over d\lambda_i^2} +{2\over \hbar} a(\lambda_i) {d\over d\lambda_i} -{M\over \hbar} \right) \prod_k \psi(\lambda_k)
$$

Since this formula holds for a basis of eigenvectors, we can ``divide'' by  $\prod_k \psi(\lambda_k)$. 
Inverting the Cauchy matrix $B_{ij}=1/(\lambda_i-\epsilon_j)$, and taking care of the order of operators
 we obtain the Hamiltonians $H_j$ in terms of the separated variables
$$
H_j = - B^{-1}_{ji} V_i, \quad B_{ij} = {1\over  \lambda_i - \epsilon_j}, \quad
V_i = {d^2\over d\lambda_i^2} +{2\over \hbar} a(\lambda_i) {d\over d\lambda_i} -{M\over \hbar} 
$$
explicitly, they are given by eqs.(\ref{hamsepa}). These Hamiltonians  are known to commute \cite{At68, ER01, BaTa03}. 
\square

To be able to work in this representation, we need  the scalar product. We set
\begin{equation}
||\Omega||^2 = \int \prod_i d\lambda_i d\bar{\lambda}_i \; W \bar{W} \;  \rho( x_1, x_2, \cdots, x_n)
\prod_i | \psi(\lambda_i)|^2 
\label{scalarproduct}
\end{equation}
where
$$
W = \prod_{i\neq j} (\lambda_i-\lambda_j)
$$
and
$$
x_i = {1\over \hbar} { \prod_j |\epsilon_i - \lambda_j|^2\over \prod_{k\neq i}(\epsilon_i - \epsilon_k)^2}
$$
The measure $\rho(x_1, x_2, \cdots, x_n)$ is determined by requiring that the Hamiltonian $H_j$ are Hermitian.

\begin{proposition}
The Hamiltonians $H_j$ are Hermitian with respect to the scalar product eq.(\ref{scalarproduct}) if
 \begin{equation}
\rho(x_1, x_2, \cdots, x_n) =  \int_0^\infty dy e^{-y} y^{M+n-\sum_i (s_i+1/2) } \prod_i
{J_{2s_i+1}(2\sqrt{y x_i}) \over x_i^{s_i+1/2} }
\label{rho}
\end{equation}
where $J_{2s_i+1}(x)$  is the Bessel function.
For $n=1$, the formula for $\rho(x)$ can be simplified  giving
$$
\rho(x) =  \partial_x^{M+1}\left[ e^{-x}  x^{M-2s}  \right]  = e^{-x} P_{M-2s}(x)
$$
where $P_{M-2s}(x)$ is a Laguerre polynomial of degree $M-2s$. 

\end{proposition}
\proof
We have to show that
\begin{equation}
 \int  \prod_k d\lambda_k d\bar{\lambda}_k 
| \psi(\lambda_k)|^2 \sum_j \left(- {d^2\over d\lambda_j^2} +{2\over \hbar} {d\over d\lambda_j} \cdot a(\lambda_j) +{M\over \hbar} \right) B^{-1}_{ij} |W|^2 \rho(x_1,\cdots,x_n)
\label{hermiHi}
\end{equation}
is real.  Now
$
B^{-1}_{ij} = \Delta^{-1} \Delta_{ji}
$
where $\Delta_{ji}$ is the minor of the element $B_{ji}$. It is clearly independent of $\lambda_j$. Hence
$$
{d\over d\lambda_j} B_{ij}^{-1} = B^{-1}_{ij} \left( {d\over d\lambda_j} - \Delta^{-1} {d\over d\lambda_j} \Delta \right)
$$
We have
$$
\Delta = {\prod_{i\neq j} (\lambda_i-\lambda_j) \prod_{i\neq j} (\epsilon_i-\epsilon_j) \over
\prod_{i,j}(\lambda_i-\epsilon_j)} =  \prod_{j\neq i}(\epsilon_i-\epsilon_j)^{-1} W \prod_{i=1}^n z_i^{-1}
$$ 
where we introduced
$$
z_i = {\prod_j(\epsilon_i - \lambda_j) \over \prod_{j\neq i}(\epsilon_i-\epsilon_j) },\quad
x_i = {z_i \bar{z}_i \over \hbar}
$$
These variables satisfy $ {d\over d\lambda_j}  z_k = B_{jk} z_k $
 so that $\Delta^{-1} {d\over d\lambda_j} \Delta  = W^{-1} {d\over d\lambda_j} W - \sum_{k} B_{jk} $
and therefore
$$
{d\over d\lambda_j} B_{ij}^{-1} |W|^2 = B_{ij}^{-1} |W|^2 \left( {d\over d\lambda_j} + \sum_k B_{jk} \right)
$$
$$
{d^2\over d\lambda_j^2} B_{ij}^{-1} |W|^2 = B_{ij}^{-1} |W|^2 \left( {d^2\over d\lambda_j^2} +2 \sum_k B_{jk}{d\over d\lambda_j} +2 \sum_k B_{jk} \sum_{l\neq k} {1\over \epsilon_k-\epsilon_l}  \right)
$$

Next, we have
$$
{d\over d\lambda_j}  \rho(x_1,\cdots,x_n)
= \sum_k B_{jk} x_k {\partial \over \partial x_k} \rho(x_1,\cdots,x_n)
$$
\begin{eqnarray*}
{d^2\over d\lambda_j^2}  \rho(x_1,\cdots,x_n)
&=& \sum_{k,l} B_{jk}B_{jl}  x_k  x_l {\partial^2 \over \partial x_k \partial x_l} \rho(x_1,\cdots,x_n)\\
&=&  \sum_{k} B_{jk}^2 x_k^2   {\partial^2 \over \partial x_k^2 } \rho(x_1,\cdots,x_n) 
 + 2 \sum_{k,l} B_{jk} {1\over \epsilon_k-\epsilon_l} x_k  x_l {\partial^2 \over \partial x_k \partial x_l} \rho(x_1,\cdots,x_n)
\end{eqnarray*}
Putting   everything together   eq.(\ref{hermiHi}) becomes
\begin{equation}
 \int \prod_l d\lambda_l d\bar{\lambda}_l 
| \psi(\lambda_l)|^2 |W|^2  \sum_j B_{ij}^{-1} \left\{ - \sum_k B_{jk}^2 x_k {\cal D}_k + \sum_k B_{jk} {\cal O}_k + {1\over \hbar}{\cal D}_0 \right\} \rho(x_1,\cdots,x_n)
\label{half}
\end{equation}
where we have defined
$$
{\cal D}_k  = x_k \partial_{x_k}^2 + 2(s_k+1) \partial_{x_k} 
\quad
{\cal D}_0 =  \sum_k x_k \partial_{x_k}  + M+n+1
$$
and
\begin{eqnarray*}
 {\cal O}_k &=& {1\over \hbar}\left(\epsilon_k - {\omega\over 2} \right) 
 - 2 \sum_{l\neq k} {1\over \epsilon_k-\epsilon_l} \Big( ( x_k  \partial_{x_k}
 + s_k + 1 ) ( x_l  \partial_{x_l}
 + s_l + 1 ) - s_k s_l -1 \Big)
 \end{eqnarray*}
The conditions on $\rho(x_1,\cdots,x_n)$ are that the sum over $j$ in  eq.(\ref{half}) should be equal to its complex conjugate.
When we perform this sum, we first get 
$$
\sum_{j,k} B_{ij}^{-1} B_{jk} {\cal O}_k\; \rho =  {\cal O}_i\; \rho
$$
which is real and gives no condition. Next we have the identities
\begin{eqnarray*}
\sum_{j} B_{ij}^{-1} &=& - z_i \\
\sum_{j} B_{ij}^{-1} B_{jk}^2&=& -{1\over \epsilon_i-\epsilon_k} {z_i\over z_k} , \quad i\neq k\\
\sum_{j} B_{ij}^{-1} B_{ji}^2&=& -{1\over z_i} - \sum_{k\neq i} {1\over \epsilon_i-\epsilon_k} {z_k\over z_i}
\end{eqnarray*}
The conditions on $\rho(x_1,\cdots,x_n)$ then read
\begin{eqnarray*}
-(z_i-\bar{z}_i) \left[ {\cal D}_i\rho + {\cal D}_0\rho \right] 
+ \sum_{k\neq i} {z_k \bar{z}_i - \bar{z}_k z_i \over \epsilon_i - \epsilon_k} 
\left[ {\cal D}_i \rho - {\cal D}_k \rho \right]  = 0
\end{eqnarray*}
Finally we find the $n$ conditions
\begin{eqnarray}
 {\cal D}_i \rho - {\cal D}_k \rho&=&0,\quad k\neq i \label{cond1} \\
  {\cal D}_i\rho + {\cal D}_0\rho &=& 0 \label{cond2}
 \end{eqnarray}
Notice that  eq.(\ref{cond2}) is independent of $i$ if the  conditions eq.(\ref{cond1}) are satisfied. A solution of eq.(\ref{cond1}) is
$$
\rho(x_1,\cdots,x_n) =  \sum_{p=0}^\infty C_p   \sum_{ q_1 + \cdots + q_n =p}  \prod_{i=1}^n  {  x_i^{q_i} \over q_i! (2s_i+1 +q_i)!} 
$$
Then eq.(\ref{cond2}) gives
$$
C_{p+1} + (M+n+p+1) C_p = 0
$$
the solution of which is
$$
C_p = (-1)^{p}  \pmatrix{M+n +p \cr p} p!   
$$
Hence we have found
 \begin{equation}
\rho(x_1, x_2, \cdots, x_n) =
 \sum_{p=0}^\infty (-1)^{p} 
 \pmatrix{M+n +p \cr p} p!   \sum_{ q_1 + \cdots + q_n =p}  \prod_{i=1}^n  {  x_i^{q_i} \over q_i! (2s_i+1 +q_i)!} 
 \label{mesurenspins}
\end{equation}
This is equivalent to eq.(\ref{rho}).
\square

\bigskip

This important formula should be further studied. In particular, for $\psi(z)$ being a Bethe state, one should be able to compute it exactly
because we know that by Gaudin formula
$$
  ||  \Omega(\mu_1, \mu_2, \cdots, \mu_M) ||^2  \simeq  \det J
 $$
where $J$ is the Jacobian matrix of Bethe's equations. This is still very mysterious.

\section{Semi-Classical limit}

The exact formula relating $Q(z)$ and $\psi(z)$, eq.(\ref{Qpsi}), allows to study the properties
of the solutions of Bethe roots $\mu_i$ in the  quasi-classical  limit $\hbar \to 0$. 
Let us set
$$
y(z)=\hbar {Q'(z)\over Q(z)}
$$
Then Baxter's equation, eq.(\ref{baxter}), becomes
\begin{equation}
\hbar\; y'(z) + y^2(z) = \Lambda(z)
\label{baxterricati}
\end{equation}
where $\Lambda(z)$ is defined in eq.(\ref{Lambda}). This is just another form of eq.(\ref{ricati}).
In the semi-classical limit eq.(\ref{baxterricati}) becomes  the equation of the spectral curve of the model
(in that limit $\hbar s_j = O(\hbar^0)$):
$$
y^2(z) = \Lambda(z)
$$
>From eq.(\ref{psi-and-q}) we deduce that 
$$
y(z) = a(z) + \sum_i {\hbar \over z-\mu_i}
$$
so that we expect in the semi-classical limit
$$
\sum_i {\hbar \over z-\mu_i} \simeq  \sqrt{\Lambda(z)} -a(z)
$$
This is a remarkable formula. It gives us the distribution of Bethe roots $\mu_i$ in the semi-classical limit,
as we now show.
Let $\sqrt{\Lambda(z)}$ be represented as a meromorphic function in the cut $z$-plane. Let us  put the cuts so that
$$
\sqrt{\Lambda(z)} = {2z \over g^2} -{\omega \over g^2} + O(z^{-1}), \quad 
  |z| \to \infty
$$
and (we neglect terms  of order $\hbar$ which do not contribute in the leading $\hbar$ approximation).
$$
 \sqrt{\Lambda(z)} = - {\hbar s_j \over z-\epsilon_j } + O(1)
,\quad z\to \epsilon_j
$$
By Cauchy theorem, we have
$$
\sqrt{\Lambda(z)} = \int_{\cal C} {dz' \over 2i\pi} {\sqrt{\Lambda(z')} \over z'-z}
$$
where $\cal C$ is composed of a big circle $C_0$ at infinity, minus small circles $C_j$ around $z=\epsilon_j$, minus 
contours $A_i$ around the cuts of $\sqrt{\Lambda(z)}$. Hence
$$
\sqrt{\Lambda(z)} = \int_{ C_0} {dz' \over 2i\pi} {\sqrt{\Lambda(z')} \over z'-z}
- \sum_j  \int_{ C_j} {dz' \over 2i\pi} {\sqrt{\Lambda(z')} \over z'-z}
- \sum_i \int_{A_i } {dz' \over 2i\pi} {\sqrt{\Lambda(z')} \over z'-z}
$$
But 
$$
 \int_{ C_0} {dz' \over 2i\pi} {\sqrt{\Lambda(z')} \over z'-z} = (\sqrt{\Lambda(z)})_+ = {2z \over g^2} -{\omega \over g^2} 
 $$
 and 
$$
 \int_{ C_j} {dz' \over 2i\pi} {\sqrt{\Lambda(z')} \over z'-z}=  \int_{ C_j} {dz' \over 2i\pi} {- \hbar s_j \over (z'-\epsilon_j)(z'-z)} = - {\hbar s_j \over z-\epsilon_j}
 $$ 
so that we arrive at
$$
\sqrt{\Lambda(z)}  - a(z) = - \sum_i \int_{A_i } {dz' \over 2i\pi} {\sqrt{\Lambda(z')} \over z'-z}
$$
and therefore we should identify
$$
\sum_i {\hbar\over z-\mu_i} =  \sum_i \int_{A_i } {dz' \over 2i\pi} {\sqrt{\Lambda(z')} \over z-z'} +  O(\hbar)
$$
Comparing both members of this formula suggests that the Bethe roots $\mu_i$ accumulate in the semiclassical limit on curves $A_i$ along which the singularities of both side should match.
To determine these curves we assume that the Bethe roots  $\mu_i$ tend to a continuous function $\mu(t)$ when $\hbar \to 0$ ( $t=\hbar\;  i$ and $i=O( \hbar^{-1})$).
$$
\sum_i {\hbar \over z-\mu_i} = \sum_i {\hbar\over z - \mu(i)} \simeq  \int  {d t \over z - \mu(t)} = 
\int_{{\cal A}} d\mu  \left({dt \over d\mu}\right) {1\over z-\mu}
$$
Here ${\cal A} = \sum A_i$. Hence, comparing with the semi-classical result, we conclude that the function $\mu(t)$ should satisfy the differential equation
\begin{equation}
{d\mu(t) \over dt} = {2i\pi\over \sqrt{\Lambda(\mu(t))}}
\label{curve}
\end{equation}
The boundary condition is that the integral curve $\mu(t)$ should start (and end !) at a branch point 
of the spectral curve $y^2 = \Lambda(z)$.  We stress that the function $\Lambda(z)$ is completely determined by the Bethe equations themselves so that these equations ``know'' the Riemann surface.

\begin{figure}[hbtp]
\begin{center}
\includegraphics[height= 22cm]{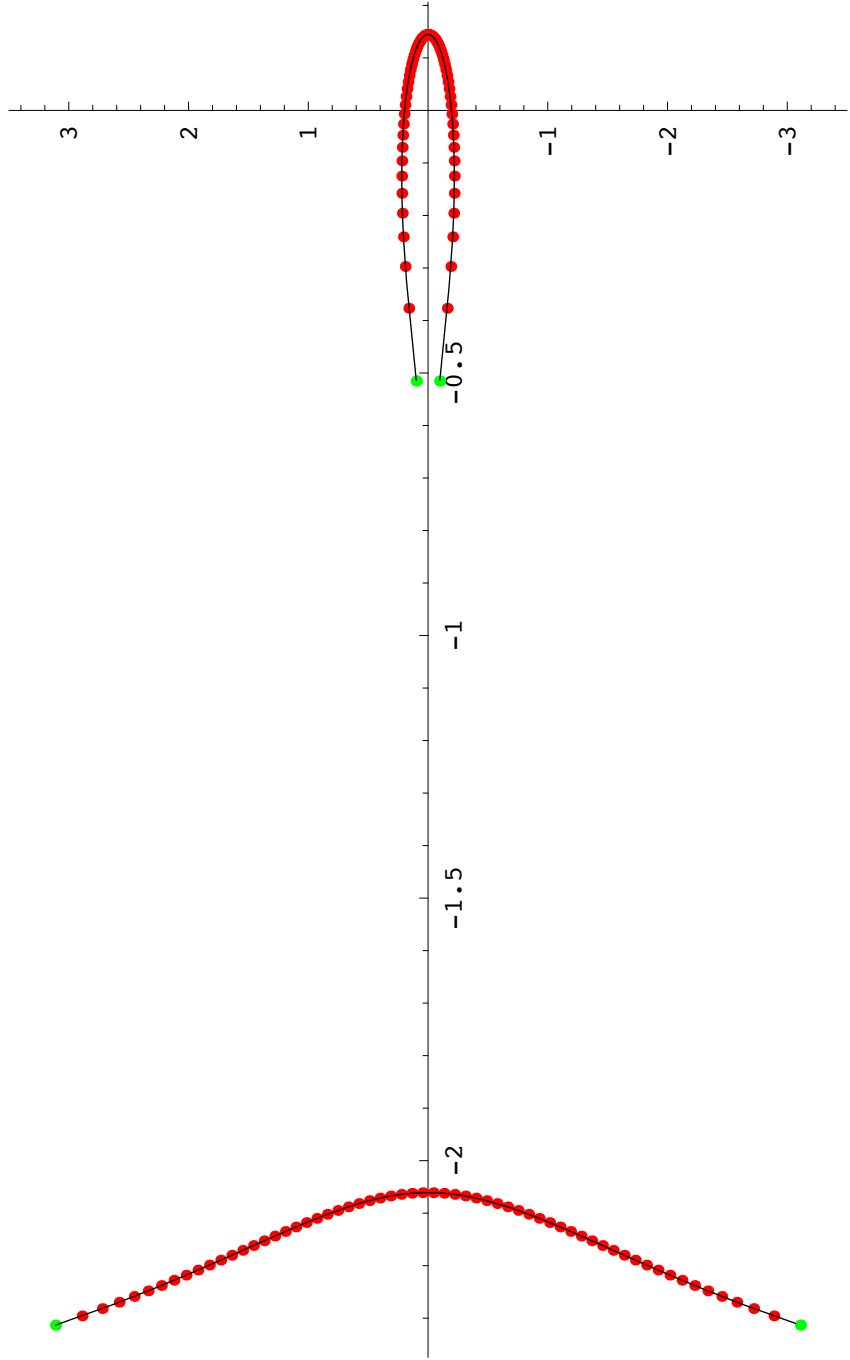} 
\caption{Red dots are the Bethe roots $\mu_i$ for the one spin system. Green dots are the branch points.
The thin black curve is the solution of eq.(\ref{curve}). ($\hbar = 1/30$, $s=1/\hbar$, $M=4/\hbar$, highest energy state).}
\label{spectacular}
\end{center}
\end{figure}

This result can be checked by numerical calculation. For simplicity, we consider the one spin-s system (n=1). A typical situation is shown in Fig.(\ref{spectacular}). The agreement is spectacular.

We can say a word on how the Bethe equations were solved. We first determine $S(\epsilon)$ by solving the polynomial equation eq.(\ref{E2s+1}) and then determine $\psi(z)$, eq.(\ref{defpsi}),
by solving eq.(\ref{eqpsi}). The Bethe roots are then obtained by solving the polynomial equation  $\psi(z)=0$.

The idea that the Bethe roots condense in the semi classical limit to form the cuts of the spectral curve
goes back to \cite{ReSmi}. It plays a very important role in the recent studies on the AdS/CFT correspondence in which it was greatly developed  \cite{Kaza04,Kaza07}. 
Eq.(\ref{curve}) however seems to be new.

\section{Appendix: The XXX spin chain}

In the Gaudin model, the Bethe equations were shown to be equivalent to a Riccati equation eq.(\ref{ricati}). Moreover this equation itself determines the parameters $S(\epsilon_j)$, i.e. the eigenvalues
of the commuting Hamiltonians. We show that this construction can be extended to the XXX spin chain.

In the case of the XXX spin chain the Bethe equations take the form (see e.g. \cite{Fa96})
\begin{equation}
\left( {\mu_j + {i\hbar \over 2} \over  \mu_j - {i\hbar\over 2}  }\right)^N = \prod_{k\neq j} { \mu_j - \mu_k +i \hbar\over 
\mu_j -\mu_k -i\hbar }
\label{BetheXXX}
\end{equation}
and the corresponding generating function for the eigenvalues of the commuting Hamiltonians  is
\begin{equation}
t(\lambda;\{\mu_j\})= \left( \lambda + {i\hbar \over 2} \right)^N \prod_k {\lambda - \mu_k -i \hbar \over \lambda - \mu_k}
+ \left( \lambda - {i\hbar \over 2} \right)^N \prod_k {\lambda - \mu_k + i\hbar  \over \lambda - \mu_k}
\label{XXXeigenval}
\end{equation}
Let us introduce the polynomial
$$
Q(\lambda) = \prod_{m=1}^M (\lambda - \mu_m)
$$
Then the Bethe equations eq.(\ref{BetheXXX}) can be rewritten as
$$
\left( \mu_k + {i\hbar \over 2} \right)^N Q(\mu_k -i\hbar ) + 
 \left( \mu_k - {i\hbar \over 2} \right)^N Q(\mu_k +i\hbar ) =0
$$
This means that the polynomial of degree $N+M$
$$
\left( \lambda + {i\hbar \over 2} \right)^N Q(\lambda -i\hbar ) + 
 \left( \lambda - {i\hbar \over 2} \right)^N Q(\lambda +i\hbar ) 
$$
is divisible by $Q(\lambda)$. Hence there exists a polynomial $t(\lambda)$ of degree $N$ such that
\begin{equation}
\left( \lambda + {i\hbar \over 2} \right)^N Q(\lambda -i\hbar ) + 
 \left( \lambda - {i\hbar \over 2} \right)^N Q(\lambda +i\hbar )  = t(\lambda) Q(\lambda)
 \label{baxterspin1/2}
\end{equation}
This is Baxter's equation.
The polynomial $t(\lambda)$ is the same as in eq.(\ref{XXXeigenval}) because that equation can be rewritten as
$$
t(\lambda;\{\mu_j\})= \left( \lambda + {i\hbar \over 2} \right)^N {Q(\lambda -i\hbar )\over Q(\lambda)} + \left( \lambda - {i\hbar \over 2} \right)^N {Q(\lambda +i\hbar ) \over Q(\lambda) }
$$
hence the coefficients of this polynomial are the eigenvalues of the set of commuting Hamiltonians.

Just as in the Gaudin model, it is interesting to introduce the Riccati version of eq.(\ref{baxterspin1/2}). We set
$$
S(\lambda) = {Q(\lambda -i\hbar ) \over Q(\lambda)}
$$
Then Baxter's equation becomes
$$
\left( \lambda + {i\hbar \over 2} \right)^N  S(\lambda) +  \left( \lambda - {i\hbar \over 2} \right)^N S^{-1}(\lambda+i\hbar )
= t(\lambda)
$$
This equation determines both $S(\lambda)$ and $t(\lambda)$.  To find the equation for $t(\lambda)$,
we expand around $\lambda= -i\hbar /2$ getting
$$
(\epsilon-i\hbar )^N S^{-1}(\epsilon + i\hbar /2) = t(\epsilon -i\hbar /2) - \epsilon^N S(\epsilon-i\hbar /2)
$$
Similarly, expanding around $\lambda = i\hbar /2$ we get
$$
(\epsilon+ i\hbar )^N S(\epsilon + i\hbar /2) = t(\epsilon + i\hbar /2) - \epsilon^N S^{-1}(\epsilon+3i\hbar /2)
$$
Multiplying the two, we find
\begin{equation}
t\left( \epsilon +{i\hbar \over 2} \right)t\left( \epsilon - {i\hbar \over 2} \right)
= (\hbar ^2+\epsilon^2)^N + O(\epsilon^N)
\label{ttspin1/2}
\end{equation}
This is a system of $N$ equations for the $N+1$ coefficients of $t(\lambda)$ which determines 
it completely  if we remember that $t(\lambda) = 2 \lambda^N + O(\lambda^{N-1})$.  Eq.(\ref{ttspin1/2})
is the characteristic equation of the commuting Hamiltonians of the XXX spin chain.

This construction can be generalized to the case of a spin-s chain.  Baxter's equation reads
\begin{equation}
\left( \lambda + {i \hbar s} \right)^N Q(\lambda -i\hbar ) + 
 \left( \lambda - {i \hbar s} \right)^N Q(\lambda +i\hbar )  = t(\lambda) Q(\lambda)
 \label{baxterspins}
\end{equation}
and the Riccati equation becomes
\begin{equation}
\left( \lambda + {i\hbar  s} \right)^N S(\lambda ) + 
 \left( \lambda - {i\hbar  s} \right)^N S^{-1}(\lambda +i\hbar )   = t(\lambda) 
\label{riccatispins}
\end{equation}
Taking $s=1$ for instance, we expand around $\lambda = i\hbar $, $\lambda =0$, $\lambda =-i\hbar $  getting
\begin{eqnarray*}
(\epsilon + 2i\hbar )^N S(\epsilon+i\hbar ) &=& t(\epsilon +i\hbar ) + O(\epsilon^N)  \\
(\epsilon + i\hbar )^N S(\epsilon)  + (\epsilon-i\hbar )^N S^{-1}(\epsilon+i\hbar )&=& t(\epsilon)  \\
(\epsilon - 2i\hbar )^N S^{-1}(\epsilon) &=& t(\epsilon -i\hbar ) + O(\epsilon^N) 
\end{eqnarray*}
from which we deduce
$$
t(\epsilon+i\hbar ) t(\epsilon) t(\epsilon-i\hbar )  = (\epsilon-i\hbar )^N (\epsilon+2i\hbar )^N t(\epsilon-i\hbar ) 
+ (\epsilon+i\hbar )^N (\epsilon-2i\hbar )^N t(\epsilon+i\hbar ) + O(\epsilon^N)
$$
Clearly for a spin-$s$, $s\geq 0$, the degree of the  equation is $2s+1$. If however $s< 0$ the equations generically do not lead to a   finite degree equation as expected.

In the semi classical limit $\hbar \to 0$, $\hbar s \to s_{cl}$, eq.(\ref{riccatispins}) tends to
$$
\left( \lambda + {i  s_{cl}} \right)^N S(\lambda ) + 
 \left( \lambda - {i  s_{cl}} \right)^N S^{-1}(\lambda  )   = t(\lambda) 
$$
which is nothing but the spectral curve of the classical spin chain.

\bigskip

{\bf Acknowledgements.} O.B.  would like to thank B. Dou\c{c}ot and T. Paul for discussions.
The paper was written during the stay of one of the authors (D.T.) at LPTHE in autumn 2006. D.T.
would like to thank the French Ambassy in Moscow  for organization of this visit.
The work of D.T. was supported by the Federal Nuclear Energy Agency of Russia, the RFBR grant 07-02-00645, and the grant of Support for the Scientific Schools 8004.2006.2.
The work of O.B. was partially supported by the European Network  ENIGMA, MRTN-CT-2004-5652 and the ANR Program GIMP, ANR-05-BLAN-0029-01.

 \end{document}